\documentstyle[12pt]{article}
\newlength{\pubnumber} 

\catcode`\@=11
\@addtoreset{equation}{section}
\def\section{\@startsection{section}{1}{\z@}{3.5ex plus 1ex minus .2ex}
 {2.3ex plus .2ex}{\large\bf}}
\def\subsection{\@startsection{subsection}{2}{\z@}{2.3ex plus .2ex}
 {2.3ex plus .2ex}{\bf}}

\begin{document}

\begin{titlepage}
\samepage{
\rightline{\tt hep-ph/0008144}
\rightline{August 2000}
\vfill
\vfill
\begin{center}
  {\Large \bf Neutrino Flavor Oscillations without Flavor
      Mixing Angles \\}
\vfill
   {\large
    Keith R. Dienes\footnote{E-mail address:  dienes@physics.arizona.edu}
    $\,$and$\,$ Ina Sarcevic\footnote{E-mail address:  ina@physics.arizona.edu}\\}
\medskip
 {\it  Department of Physics, University of Arizona, Tucson, AZ  85721  USA\\}
\end{center}
\vfill
\vfill
\begin{abstract}
  {\rm
     We demonstrate that sizable neutrino flavor 
     oscillations can be generated in a model with large extra spacetime
      dimensions even if the physics on the brane is
     flavor-diagonal, the bulk neutrino theory is flavor-neutral,
     and the brane/bulk couplings are flavor-blind.
        This is thus a compact model for 
      addressing neutrino flavor oscillations in higher dimensions.
     We also discuss several phenomenological aspects of 
     the ``bulk-mediated'' neutrino oscillations inherent in this model,
      and show that this model contains some potentially important 
        new phenomenological features
     in the limit of large brane/bulk coupling.  }
\end{abstract}
\vfill
\vfill}
\end{titlepage}

\setcounter{footnote}{0}

\newcommand{\newc}{\newcommand}

\newc{\gsim}{\lower.7ex\hbox{$\;\stackrel{\textstyle>}{\sim}\;$}}
\newc{\lsim}{\lower.7ex\hbox{$\;\stackrel{\textstyle<}{\sim}\;$}}

\def\beq{\begin{equation}}
\def\eeq{\end{equation}}
\def\beqn{\begin{eqnarray}}
\def\eeqn{\end{eqnarray}}
\def\dnot#1{{ \not{\!\! {#1}} }}  
\def\half{{\textstyle{1\over 2}}}
\def\ie{{\it i.e.}\/}
\def\eg{{\it e.g.}\/}
\def\ket#1{{|{#1}\rangle}}

\hyphenation{su-per-sym-met-ric non-su-per-sym-met-ric}
\hyphenation{space-time-super-sym-met-ric}
\hyphenation{mod-u-lar mod-u-lar--in-var-i-ant}


\def\inbar{\,\vrule height1.5ex width.4pt depth0pt}

\def\IC{\relax\hbox{$\inbar\kern-.3em{\rm C}$}}
\def\IQ{\relax\hbox{$\inbar\kern-.3em{\rm Q}$}}
\def\IR{\relax{\rm I\kern-.18em R}}
 \font\cmss=cmss10 \font\cmsss=cmss10 at 7pt
\def\IZ{\relax\ifmmode\mathchoice
 {\hbox{\cmss Z\kern-.4em Z}}{\hbox{\cmss Z\kern-.4em Z}}
 {\lower.9pt\hbox{\cmsss Z\kern-.4em Z}}
 {\lower1.2pt\hbox{\cmsss Z\kern-.4em Z}}\else{\cmss Z\kern-.4em Z}\fi}

\def\NPB#1#2#3{{\it Nucl.\ Phys.}\/ {\bf B#1} (19#2) #3}
\def\PLB#1#2#3{{\it Phys.\ Lett.}\/ {\bf B#1} (19#2) #3}
\def\PRD#1#2#3{{\it Phys.\ Rev.}\/ {\bf D#1} (19#2) #3}
\def\PRL#1#2#3{{\it Phys.\ Rev.\ Lett.}\/ {\bf #1} (19#2) #3}
\def\PRT#1#2#3{{\it Phys.\ Rep.}\/ {\bf#1} (19#2) #3}
\def\CMP#1#2#3{{\it Commun.\ Math.\ Phys.}\/ {\bf#1} (19#2) #3}
\def\MODA#1#2#3{{\it Mod.\ Phys.\ Lett.}\/ {\bf A#1} (19#2) #3}
\def\IJMP#1#2#3{{\it Int.\ J.\ Mod.\ Phys.}\/ {\bf A#1} (19#2) #3}
\def\NUVC#1#2#3{{\it Nuovo Cimento}\/ {\bf #1A} (#2) #3}
\def\etal{{\it et al.\/}}

\long\def\@caption#1[#2]#3{\par\addcontentsline{\csname
  ext@#1\endcsname}{#1}{\protect\numberline{\csname
  the#1\endcsname}{\ignorespaces #2}}\begingroup
    \small
    \@parboxrestore
    \@makecaption{\csname fnum@#1\endcsname}{\ignorespaces #3}\par
  \endgroup}
\catcode`@=12

\input epsf


Over the past few years, 
experimental evidence 
that neutrinos have mass and undergo oscillations has been 
pointing towards the existence of new physics beyond the Standard Model.  
Recent SuperK data on solar neutrinos~\cite{Suzuki} 
and atmospheric neutrinos~\cite{Sobel}, 
combined with the LSND accelerator measurements~\cite{Mills}, 
has provided a new challenge for our understanding of 
the lepton sector of the Standard Model.  

In some sense, the observation of lepton mixing restores
a parallel structure between the quark and lepton sectors
of the Standard Model.
However, although the quark sector of the Standard Model
is characterized by relatively small flavor mixing angles,
recent data suggest that the lepton sector experiences
large flavor mixings.  
Explaining the origin of these large mixings thus becomes
a theoretical challenge.
One important observation in this regard is the
fact that right-handed quarks carry both color and
electromagnetic charge, while 
right-handed neutrinos are neutral under all
Standard-Model gauge symmetries.  
Thus,  it is plausible that right-handed
neutrinos have completely different origins 
than their quark counterparts.

One interesting possibility~\cite{DDGneut,ADDneut}  
is that the right-handed neutrino may be 
realized as a closed- (rather than open-) string state, and hence may
propagate in the higher-dimensional bulk corresponding to 
a large extra spacetime dimension.
The possibility of having large extra spacetime dimensions
has been proposed~\cite{majorpapers}
as a method of lowering the fundamental high-energy scales
of physics such as the GUT, Planck, and string scales.
Allowing the right-handed neutrino to feel these extra
spacetime dimensions would
have profound consequences for neutrino physics;  indeed,
it might then be possible to use neutrinos
as a method of probing the physics of
large extra dimensions.  
Various phenomenological aspects of higher-dimensional neutrino physics
have been explored in Refs.~\cite{DDGneut,ADDneut,others}.

One key issue concerns the embedding of flavor into such 
higher-dimensional scenarios.
All flavor models in the literature thus far  
introduce one bulk neutrino for each of the 
three brane neutrinos, thereby extending flavor into the 
bulk.  However, in such cases, 
the brane/bulk couplings become arbitrary $3\times 3$ mixing matrices 
whose parameters are undetermined.  
Moreover, the three bulk neutrinos
can in principle correspond to different 
extra spacetime dimensions with
different radii.  Thus, one obtains a scenario 
with many undetermined parameters governing neutrino masses
and mixing angles.
 
By contrast, we shall here introduce 
a ``compact'' model in which
only one bulk neutrino is required.
In other words,
we shall consider flavor
to be a property internal to the Standard Model, one which is restricted
to the brane and which therefore does not extend
into the bulk.
Furthermore we shall assume flavor-blind couplings
between our brane and bulk fields.  Thus, in this model,
only the physics
directly on the brane is flavor-sensitive.
Moreover, for simplicity we shall also assume that our brane
theory is completely flavor-diagonal. 
Despite these facts, we shall see that neutrino flavor oscillations
can still arise.

We begin by briefly describing the model.
On the brane, we introduce 
three left-handed neutrinos $\nu_i$ ($i=1,2,3$);
these are our flavor eigenstates.
We assume that these left-handed neutrinos have 
corresponding Majorana masses $m_i$ on the brane,
and we shall take 
the $m_i$ to be unequal.  It is in this way that
we shall distinguish between the different flavors on the brane.
Note that for our purposes, we shall simply consider
these Majorana masses to be arbitrary input parameters,
and we shall not speculate on their origins or sizes.
Despite the presence of these masses, however, we shall not
introduce any explicit flavor mixings between the
left-handed neutrinos on the brane.   
Thus, our theory on the brane will be completely flavor-diagonal.

Our bulk theory will be identical to that considered in Ref.~\cite{DDGneut}.
Specifically, we shall consider 
a single five-dimensional Dirac fermion $\Psi$, which in the Weyl 
basis can be decomposed
into two two-component spinors:  $\Psi = (\psi_1,\bar\psi_2)^T$.
This Dirac fermion does not carry any flavor indices, and is therefore
completely flavor-neutral.
We impose the orbifold relations
$\psi_{1,2}(-y)=\pm \psi_{1,2}(y)$
where $y$ is the coordinate of the fifth dimension.
In order to have a diagonal mass matrix, 
we also introduce the linear combinations
$N^{(n)}\equiv(\psi_1^{(n)}+\psi_2^{(n)})/\sqrt{2}$
and
$M^{(n)}\equiv(\psi_1^{(n)}-\psi_2^{(n)})/\sqrt{2}$
for all $n>0$.
Assuming that the brane is located at the orbifold fixed point $y=0$,
we see that $\psi_2$ vanishes on the brane. 
The most natural brane/bulk coupling is therefore simply between 
$\nu_i$ and $\psi_1$.
Although in principle each left-handed neutrino $\nu_i$ on the brane
can have a different coupling to $\psi_1$, 
we shall consider a simple model in which this brane/bulk
coupling, labeled by $\hat m$,
 is completely flavor-universal. 

Given these assumptions, our 
Lagrangian has the form
\beqn
  {\cal L}_{\rm brane} &=& \int d^4 x ~ 
         \sum_{i=1}^3 \, \biggl\lbrace
        {\bar\nu}_i i{\bar\sigma}^\mu D_\mu \nu_i 
        +  m_i (\nu_i\nu_i + {\rm h.c.}) \biggr\rbrace
                                 \nonumber\\
  {\cal L}_{\rm bulk} &=& \int d^{4} x \,dy ~M_s\, \biggl\lbrace
     {\bar\psi}_1 i{\bar\sigma}^\mu \partial_\mu \psi_1
       +{\bar\psi}_2 i{\bar\sigma}^\mu \partial_\mu \psi_2\biggr\rbrace 
                                 \nonumber\\
  {\cal L}_{\rm coupling} &=& \int d^4 x
         \sum_{i=1}^3\, ({\hat m} \nu_i \psi_1|_{y=0} + {\rm h.c.})~.
\label{klag}
\eeqn
Here $M_s$ is the
mass scale of the higher-dimensional fundamental
theory.  
By compactifying the Lagrangian (\ref{klag}) down to four dimensions
we obtain 
\beqn
  {\cal L} &=& \int d^4 x ~ \Biggl\lbrace
    \sum_{i=1}^3 \, {\bar\nu}_i i{\bar\sigma}^\mu D_\mu \nu_i
     + {\bar \psi}_1^{(0)} i{\bar\sigma}^\mu\partial_\mu \psi_1^{(0)}
     + \sum_{n=1}^\infty \left(
     {\bar N}^{(n)} i{\bar\sigma}^\mu\partial_\mu N^{(n)}
     +{\bar M}^{(n)} i{\bar\sigma}^\mu\partial_\mu M^{(n)} \right) \nonumber\\
    && ~~~~+~  
      \biggl\lbrace \sum_{i=1}^3 \, m_i \nu_i \nu_i  
    + \half \sum_{n=1}^\infty \, \left\lbrack
         \left({n\over R}\right) N^{(n)} N^{(n)}
         - \left({n\over R}\right) M^{(n)} M^{(n)} \right\rbrack
            \nonumber\\
    && ~~~~+~ m \sum_{i=1}^3 \nu_i \left( \psi_1^{(0)} +
           \sum_{n=1}^\infty N^{(n)}
           + \sum_{n=1}^\infty M^{(n)} \right) 
           ~+~ {\rm h.c.}\biggr\rbrace\Biggr\rbrace~
\label{tglag}
\eeqn
where $m\equiv \hat m/\sqrt{2\pi M_sR}$ is the volume-suppressed brane/bulk
coupling resulting from the rescaling of the 
individual $\psi_1^{(0)}$, $N^{(n)}$, and $M^{(n)}$ Kaluza-Klein modes.  

Given the Lagrangian (\ref{tglag}), we see that
the Standard-Model flavor-eigenstate neutrinos
$\nu_i$ will mix with the entire tower of Kaluza-Klein states
of the higher-dimensional $\Psi$ field, even though they do not mix
directly with each other.
Defining
\beq
        {\cal N}^T ~\equiv~ (\nu_1, \nu_2, \nu_3, \psi_1^{(0)},
                  N^{(1)}, M^{(1)},
                  N^{(2)}, M^{(2)}, ...)~,
\label{calNdef}
\eeq
we see that the mass terms in the Lagrangian (\ref{tglag})
take the form $\half({\cal N}^T {\cal M} {\cal N}+{\rm h.c.})$
where ${\cal M}$ takes the symmetric form
\beq
      {\cal M} ~=~ \pmatrix{
         m_1 &  0  &  0  &  m   &   m  &   m   &   m  &  m  & \ldots \cr
         0 &  m_2  &  0  &  m   &   m  &   m   &   m  &  m  & \ldots \cr
         0 &  0  &  m_3  &  m   &   m  &   m   &   m  &  m  & \ldots \cr
         m & m & m &  0  &   0  &   0  &   0  &  0  & \ldots \cr
         m & m & m &   0   &   1/R  &   0  &   0  &  0  & \ldots \cr
         m & m & m &   0   &   0  &   -1/R  &   0  &  0  & \ldots \cr
         m & m & m &   0   &   0  &   0  &  2/R  &  0  & \ldots \cr
         m & m & m &   0   &   0  &   0  &   0 & -2/R    & \ldots \cr
        \vdots & \vdots & \vdots & \vdots & \vdots & \vdots & \vdots & 
        \vdots & \ddots \cr }~.
\label{newmatrix}
\eeq

In this model, the bulk theory is flavor-neutral and the
brane/bulk couplings are flavor-blind.
Moreover, the theory on the brane is flavor-diagonal.
Nevertheless, it is immediately apparent the three brane neutrinos will 
undergo flavor oscillations
as a result of their indirect mixings with the bulk Kaluza-Klein neutrinos.
In order to demonstrate this explicitly, we can determine
the eigenvalues and eigenvectors of this mass matrix.  
It turns out that the eigenvalues $\lambda$
of the matrix (\ref{newmatrix})  
are given exactly as the solutions to the transcendental equation
\beq
     \tan \pi \lambda R ~=~ \pi m^2 R \,\sum_{i=1}^3\, {1\over \lambda-m_i}~.
\label{eigequation}
\eeq
For each solution $\lambda$ to (\ref{eigequation}), the corresponding
mass eigenstate $\ket{\tilde \nu_\lambda}$ is then exactly given by
\beqn
   \ket{\tilde \nu_\lambda} &=& 
     {1\over \sqrt{N_\lambda}}\Biggl\lbrack
           \left(\sum_{j=1}^3 {m\over \lambda-m_j}\right)^{-1}\,
     \, \sum_{i=1}^3  \, {\lambda\over \lambda-m_i} \,\ket{\nu_i}   
           ~+~ \ket{\psi_1^{(0)}} \nonumber\\ 
         &&~~~~~~~+~ \sum_{k=1}^\infty \,
         {\lambda\over \lambda - k/R}\,\ket{N^{(k)}}
         ~+~ \sum_{k=1}^\infty \,
         {\lambda\over \lambda + k/R}\,\ket{M^{(k)}}
       \Biggr\rbrack~
\label{eigenstate}
\eeqn
where $N_\lambda$ is an overall normalization constant. 

Note that the different flavor eigenstates $\ket{\nu_i}$
consist of different linear combinations of the different
mass eigenstates $\ket{\tilde \nu_\lambda}$ for each
$\lambda$.
Thus, the different flavor eigenstates will experience a
relative oscillation with each other. 
This oscillation is entirely ``bulk-mediated''
in the sense that there are no explicit flavor mixings 
on the brane;  it is
the presence of the higher-dimensional bulk which is completely  
responsible for inducing the flavor oscillations on the brane.
Note that these oscillations are similar in spirit to the
so-called ``indirect'' four-dimensional neutrino oscillations discussed
in Ref.~\cite{indirect}, except that here we have an infinite
tower of higher-dimensional Kaluza-Klein states mediating our flavor oscillations
and we have no explicit mixing angles on the brane.

We shall now discuss several phenomenological features
of the ``bulk-mediated'' flavor oscillations inherent in this model.
In general, we are interested in the probabilities $P_{i\to j}(t)$
that $\ket{\nu_i}$ oscillates into $\ket{\nu_j}$ 
as a function of time $t$.
For $i=j$, this refers to 
flavor preservation on the brane,
while for $i\not= j$ this refers to 
flavor conversion on the brane.
Note that in general, 
$P_{i\to j} = P_{j\to i}$ for all $(i,j)$.
For phenomenological purposes, we are 
mostly interested in the probabilities
$P_{1\to 1}(t)$, $P_{2\to 2}(t)$,
and $P_{2\to 1}(t)$ where the subscripts $(1,2,3)$ 
signify $(\nu_e,\nu_\mu,\nu_\tau)$ respectively. 
These are ultimately the three probabilities which are relevant
for addressing constraints from
solar, atmospheric, and long-baseline neutrino experiments.

In the following, we shall rescale all of our mass 
variables $\lbrace m_i, m\rbrace$ by $R$
so that these variables henceforth correspond to dimensionless quantities.
Thus, our model has only  four free dimensionless input parameters,
$\lbrace m, m_1, m_2, m_3\rbrace$, while the radius $R$ serves as
an overall length scale. 
These parameters ultimately determine not only
the masses of the physical neutrinos, but also their relative effective
mixing angles.
We shall also define a dimensionless time variable, 
$ \tilde t \equiv t/ (2 p R^2) \approx (1/ 2 R^2) (L/E)$,
where $L\approx ct$ is the spatial
distance between the locations of neutrino production and neutrino
detection, and $E\approx pc$ is the neutrino energy. 
We can then plot our probabilities $P_{i\to j}(t)$ as 
functions of $\tilde t$.

\begin{figure}[ht]
\centerline{ 
      \epsfxsize 2.0 truein \epsfbox {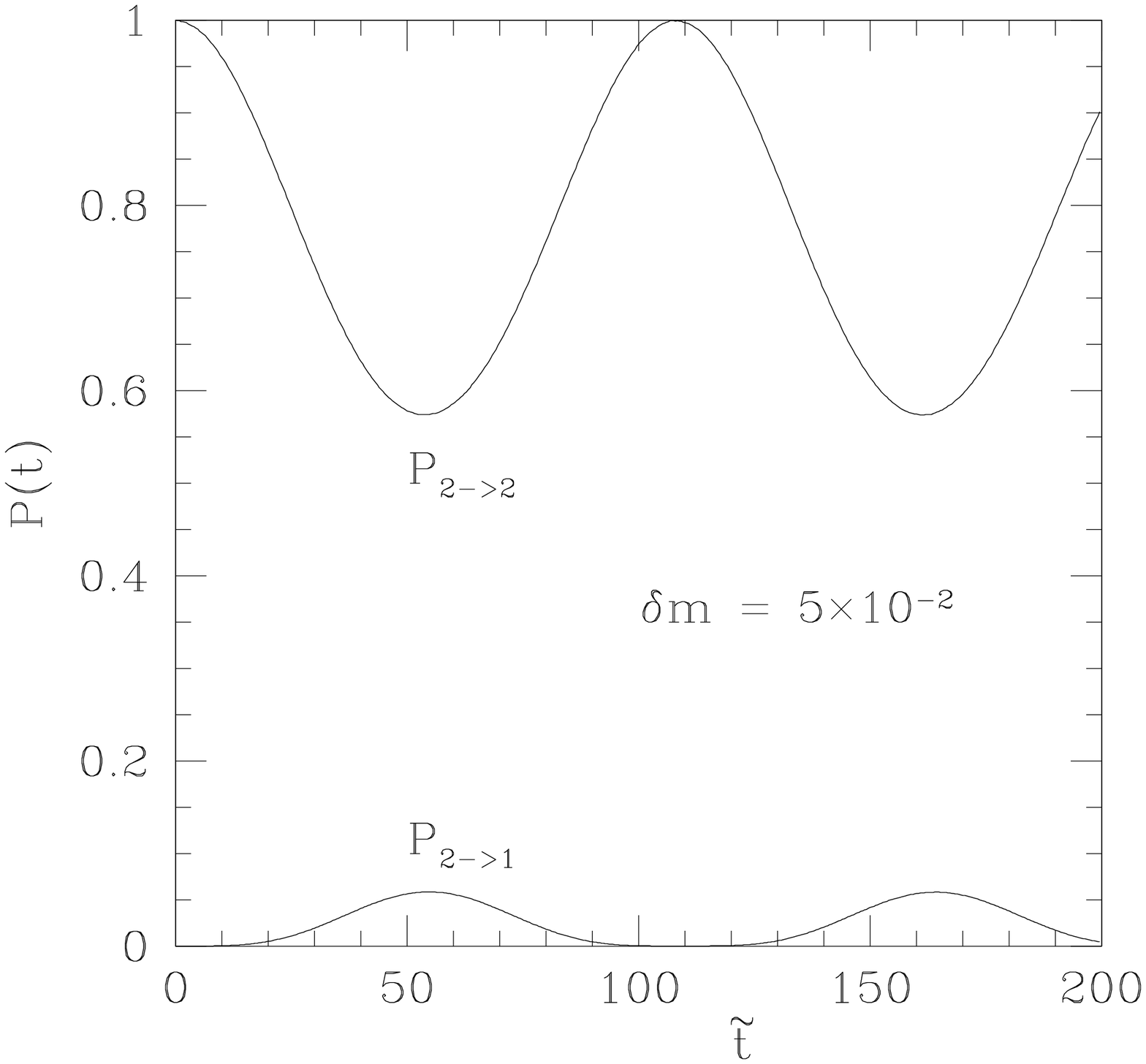}
      \epsfxsize 2.0 truein \epsfbox {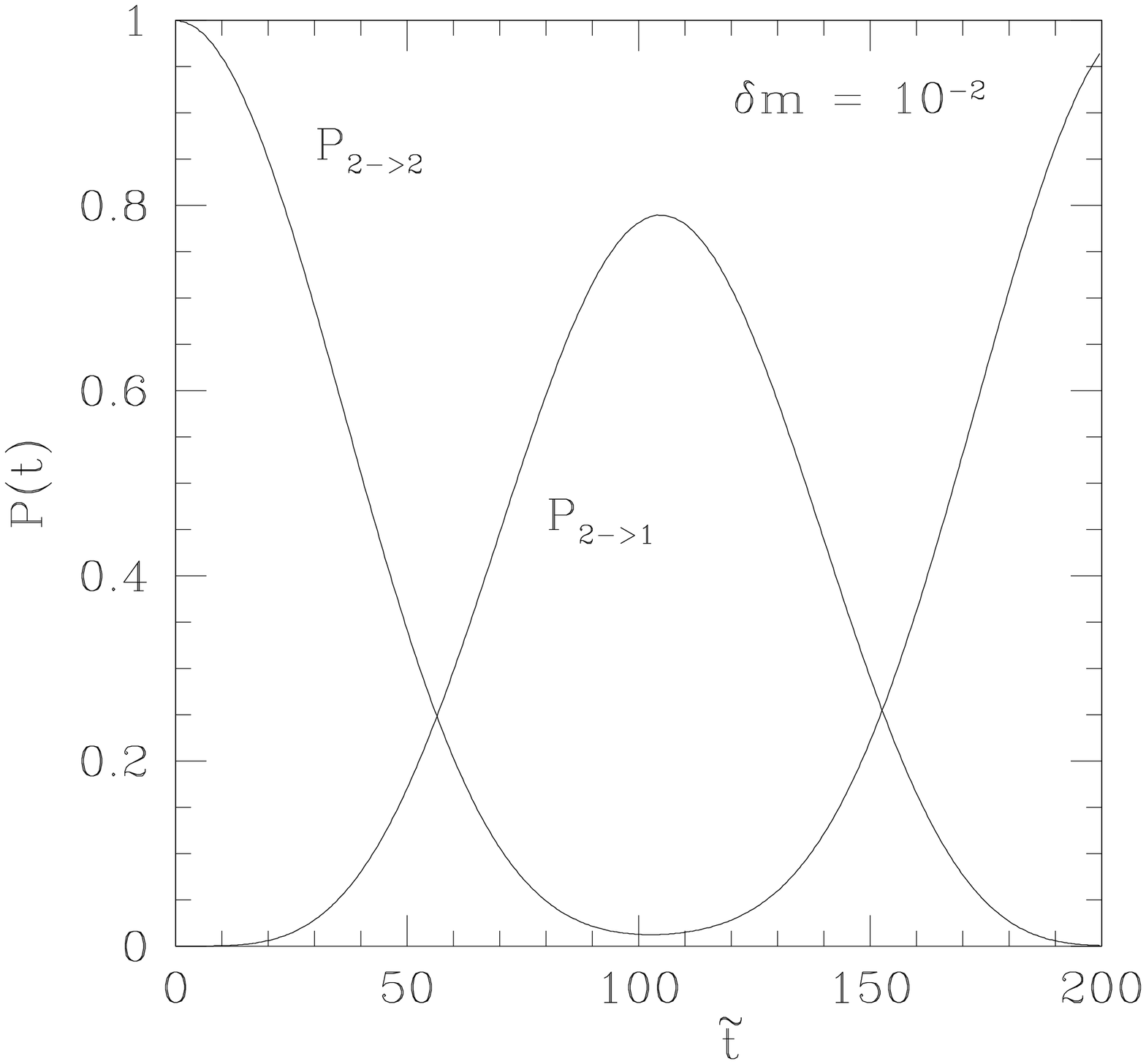}
      \epsfxsize 2.0 truein \epsfbox {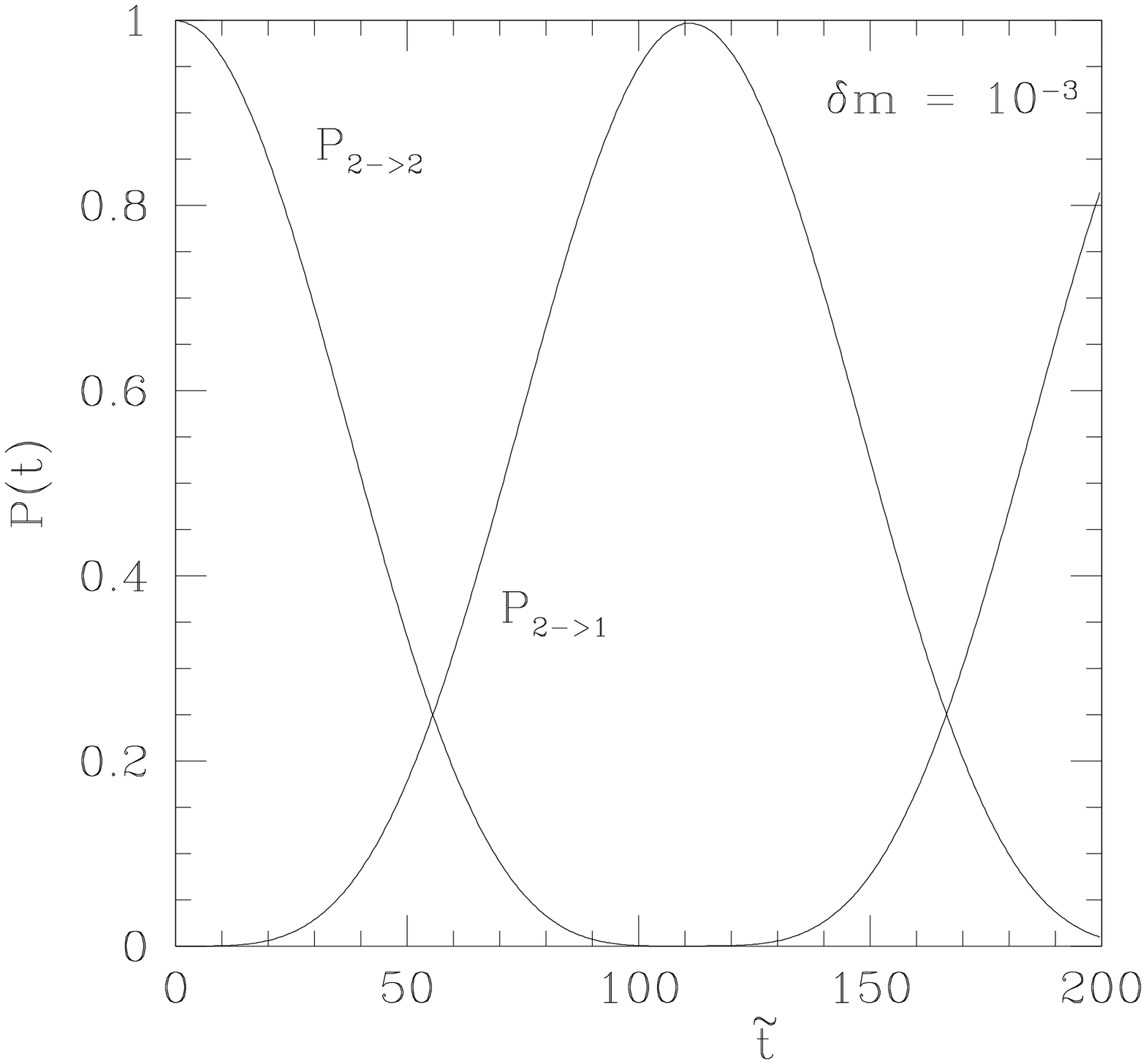}
 }
\caption{ Bulk-mediated neutrino flavor oscillations with $m=0.01$
       and $m_{1,2}= 1\mp \delta m/2$, for various $\delta m$.
         As discussed in the text, these flavor oscillations 
         are essentially ``maximal'' even though all mixing angles
         on the brane are vanishing.  }
\label{firstfig}
\end{figure}

Let us first consider the case when the 
brane/bulk coupling is extremely small, \ie, $m\ll 1$.
In such cases, the mass matrix (\ref{newmatrix})
is nearly diagonal, and 
there is relatively little mixing between
the brane neutrinos and the bulk neutrinos.
However, 
we find that large flavor oscillations can nevertheless be achieved
for certain values of the input parameters $\lbrace m, m_i\rbrace$.
As an explicit example, let us consider the case
$m=0.01$, $m_{1,2}=1\mp \delta m/2$, and $m_3=5$, 
with $\delta m = m_2-m_1$ free to vary.  
The resulting neutrino oscillations are shown in Fig.~\ref{firstfig}. 
Note that as $\delta m \to 0$, the resulting flavor oscillation becomes effectively 
maximal in the sense that we achieve full flavor conversion.
In achieving this large flavor oscillation,
we are exploiting a resonance
between two approximately degenerate brane neutrinos $\ket{\nu_{1,2}}$ 
and the first excited Kaluza-Klein bulk neutrino $\ket{N^{(1)}}$,
so that large effective flavor oscillations between the brane
neutrinos are mediated indirectly through their 
oscillations with the bulk neutrino.
Nevertheless, no mixing angles on the brane are required.

Even though this flavor mixing appears to resemble a simple two-state
oscillation in the limit $\delta m \ll 1$, 
the time-averaged probabilities for each flavor are not 1/2, but 3/8.  
This reflects the fact that we are
dealing with a three-state oscillation in which
1/4 of the total neutrino probability
has been lost to the bulk neutrino $\ket{N^{(1)}}$, and is
hence ``sterile'' from the point of view of flavor on the brane.

\begin{figure}
\centerline{ \epsfxsize 3.0 truein \epsfbox {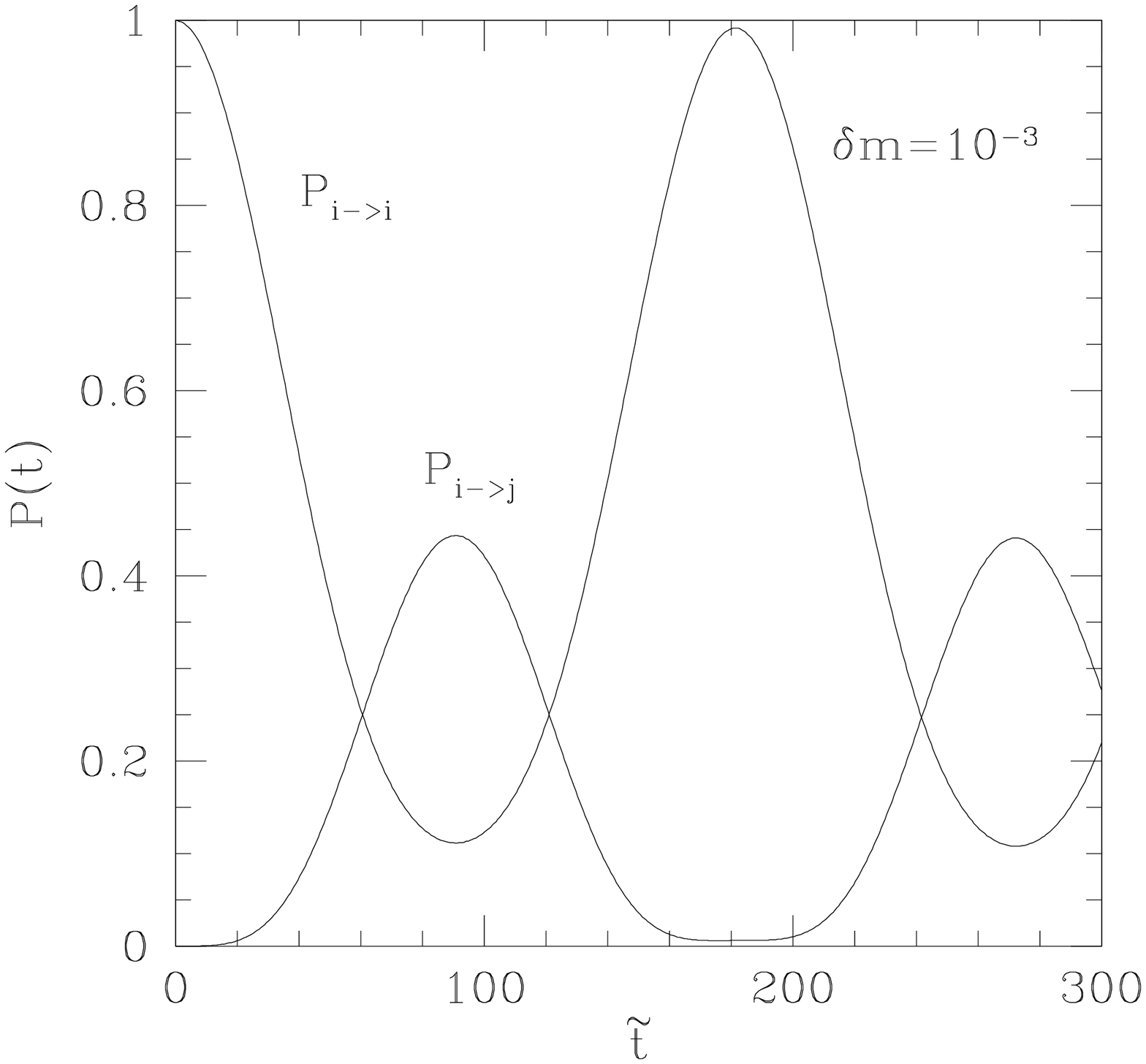}}
\centerline{
      \epsfxsize 2.0 truein \epsfbox {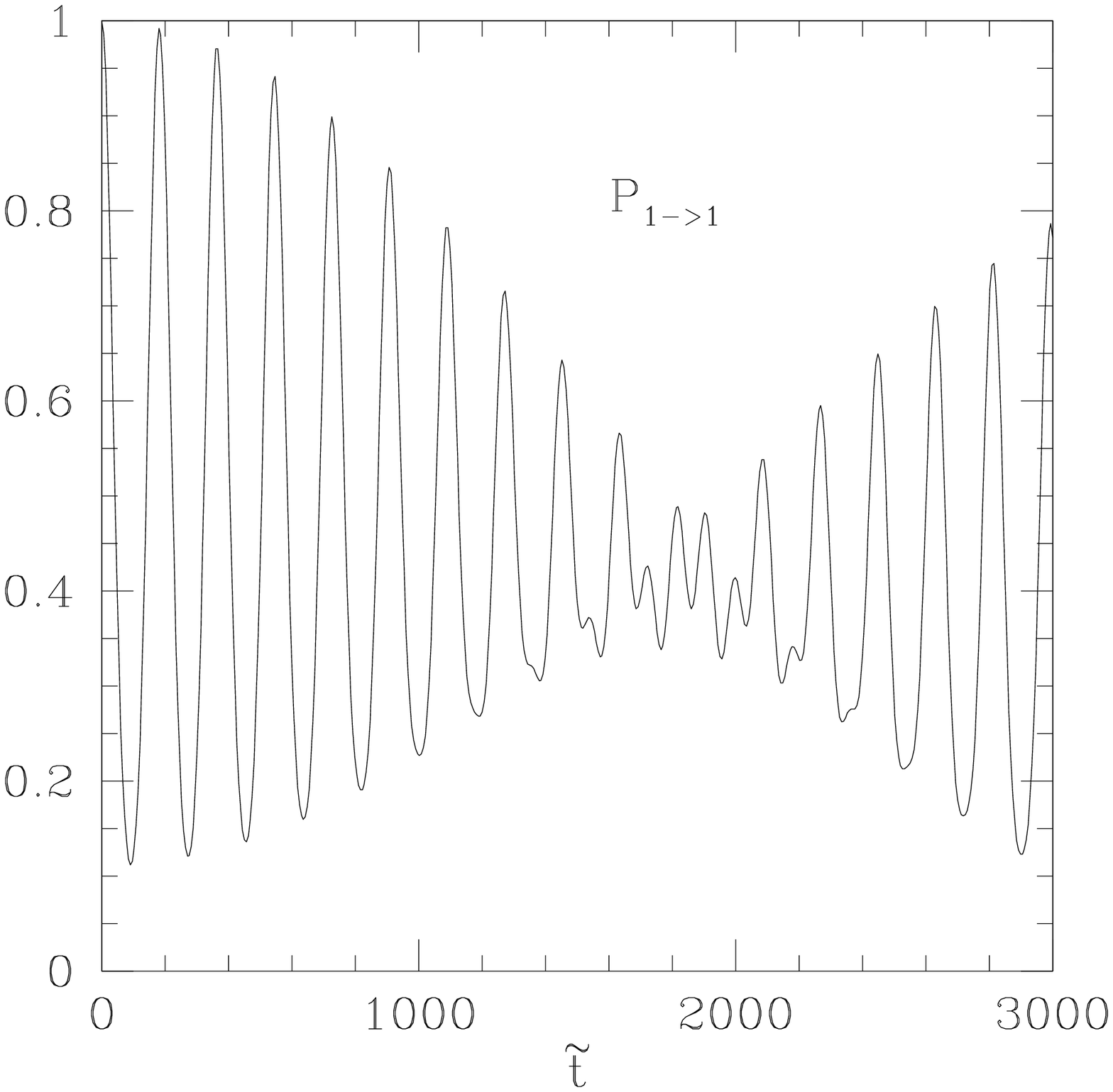}
      \epsfxsize 2.0 truein \epsfbox {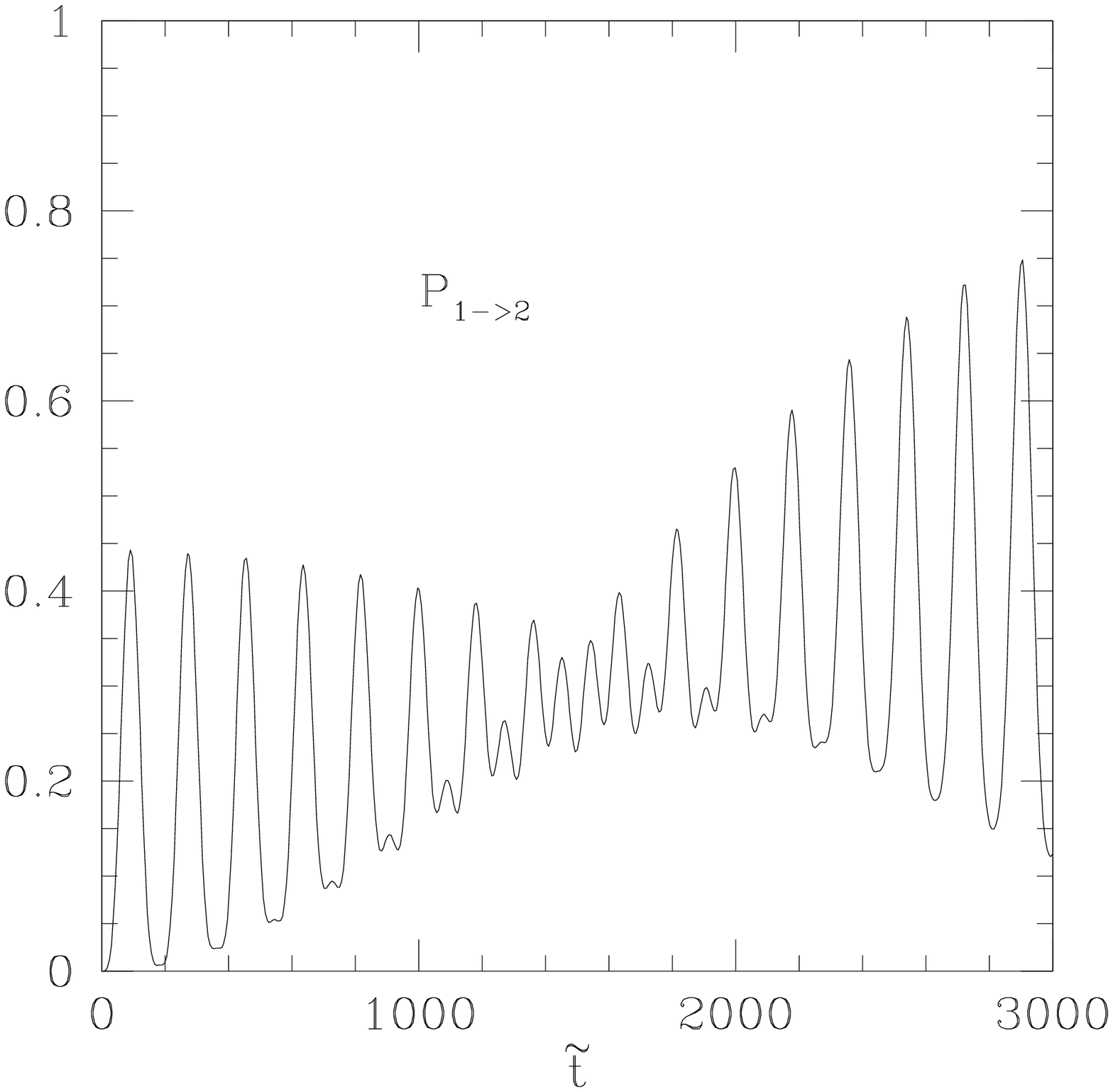}
      \epsfxsize 2.0 truein \epsfbox {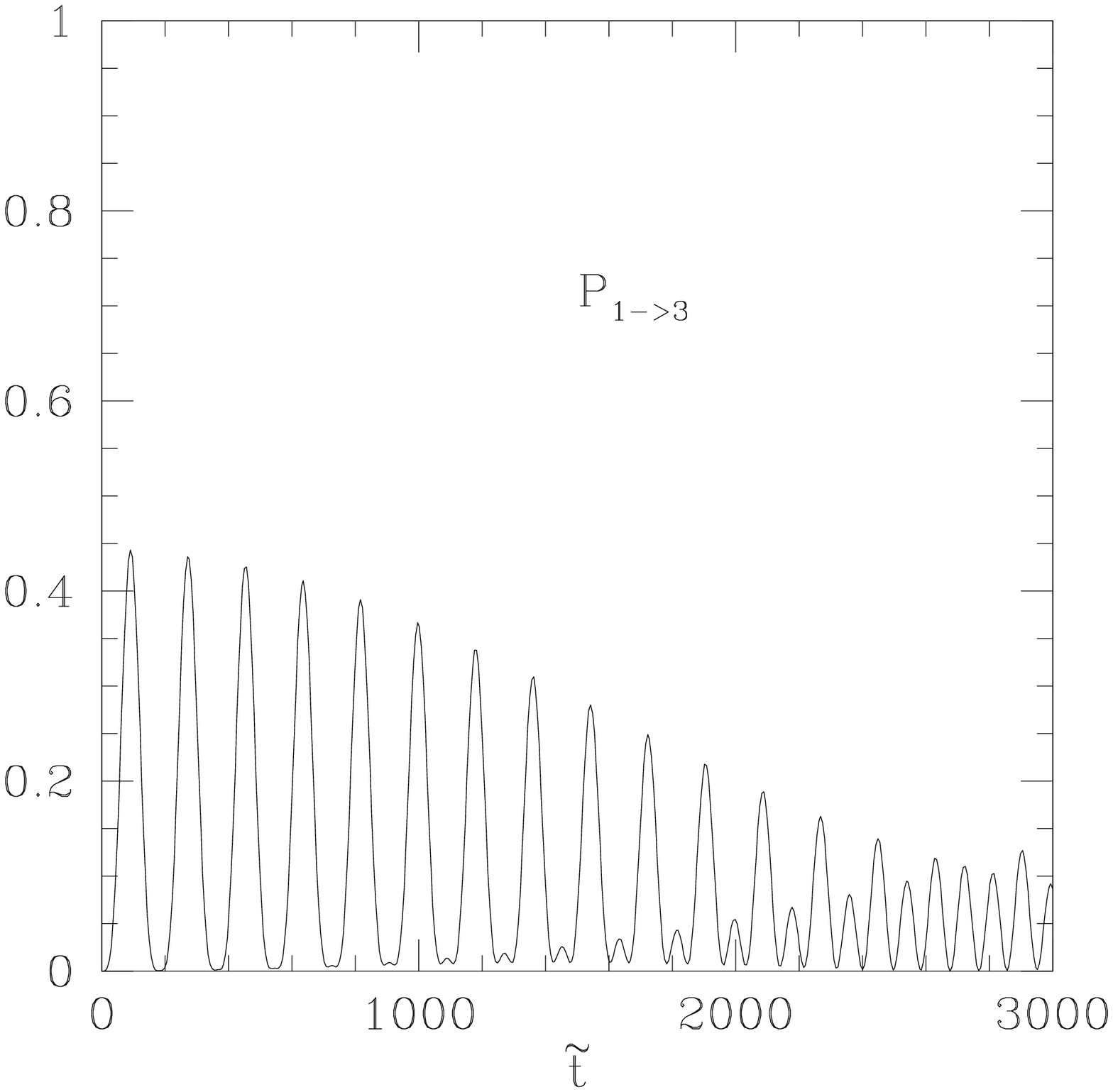}
 }
\centerline{
      \epsfxsize 2.0 truein \epsfbox {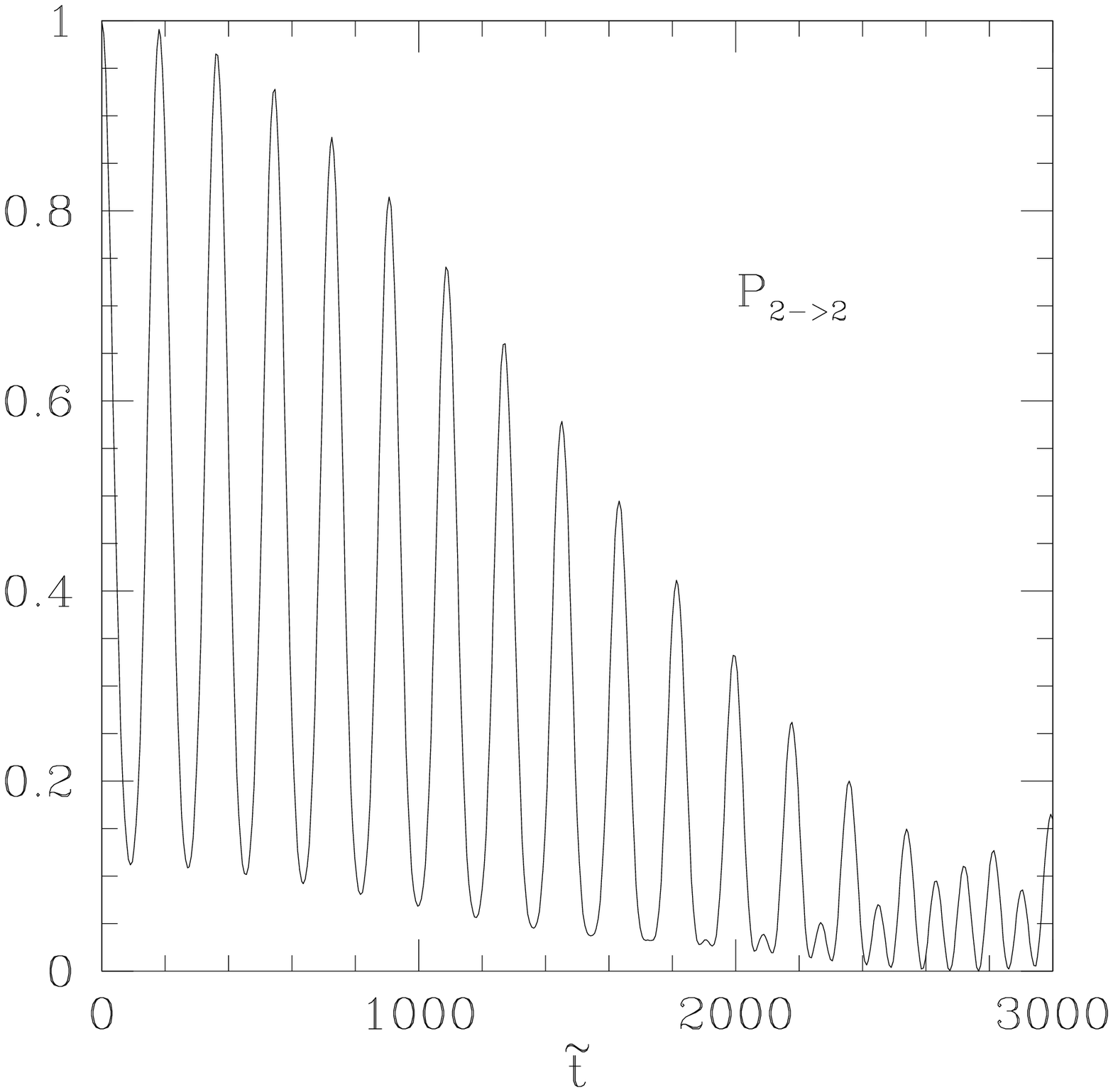}
      \epsfxsize 2.0 truein \epsfbox {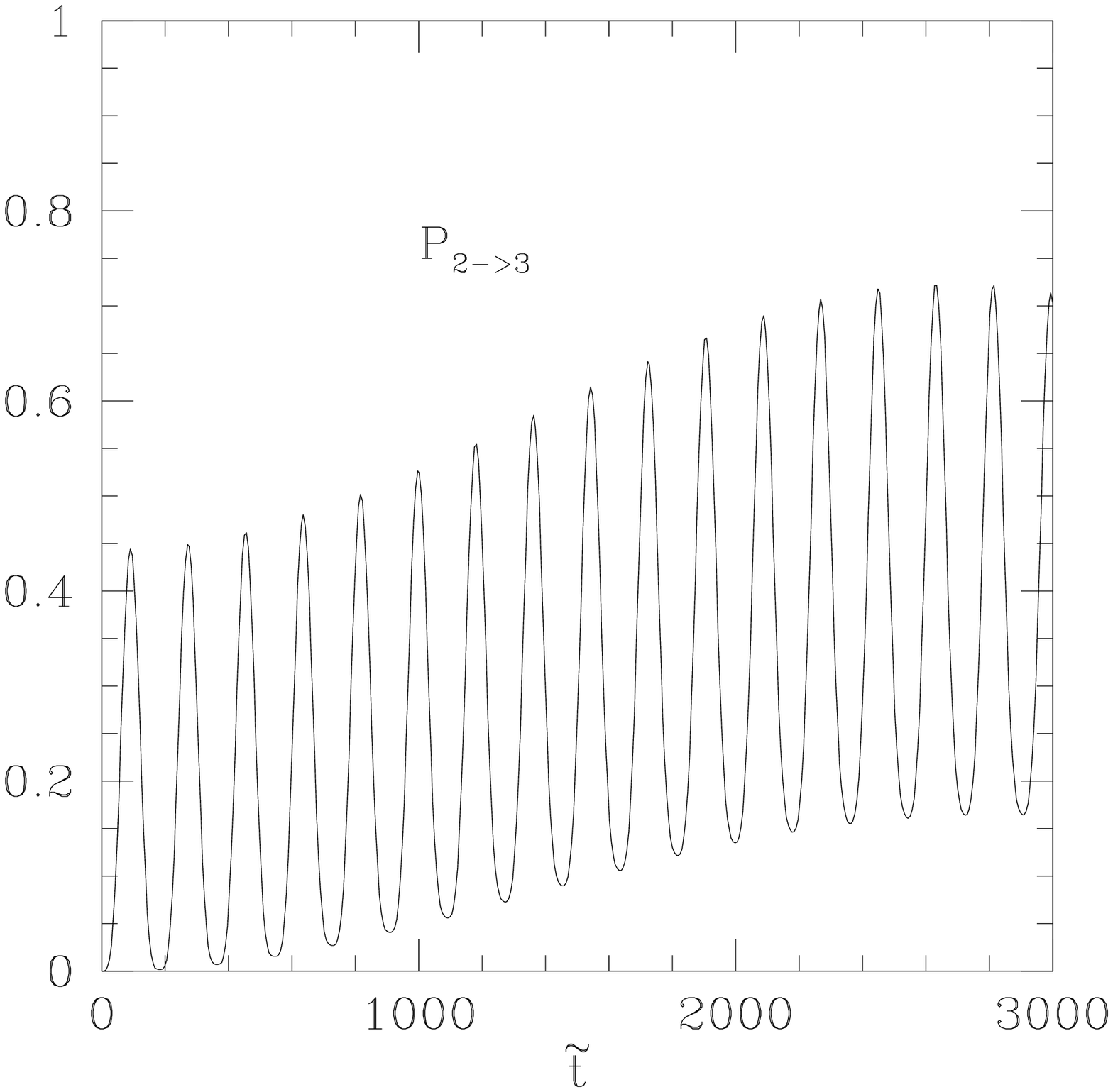}
      \epsfxsize 2.0 truein \epsfbox {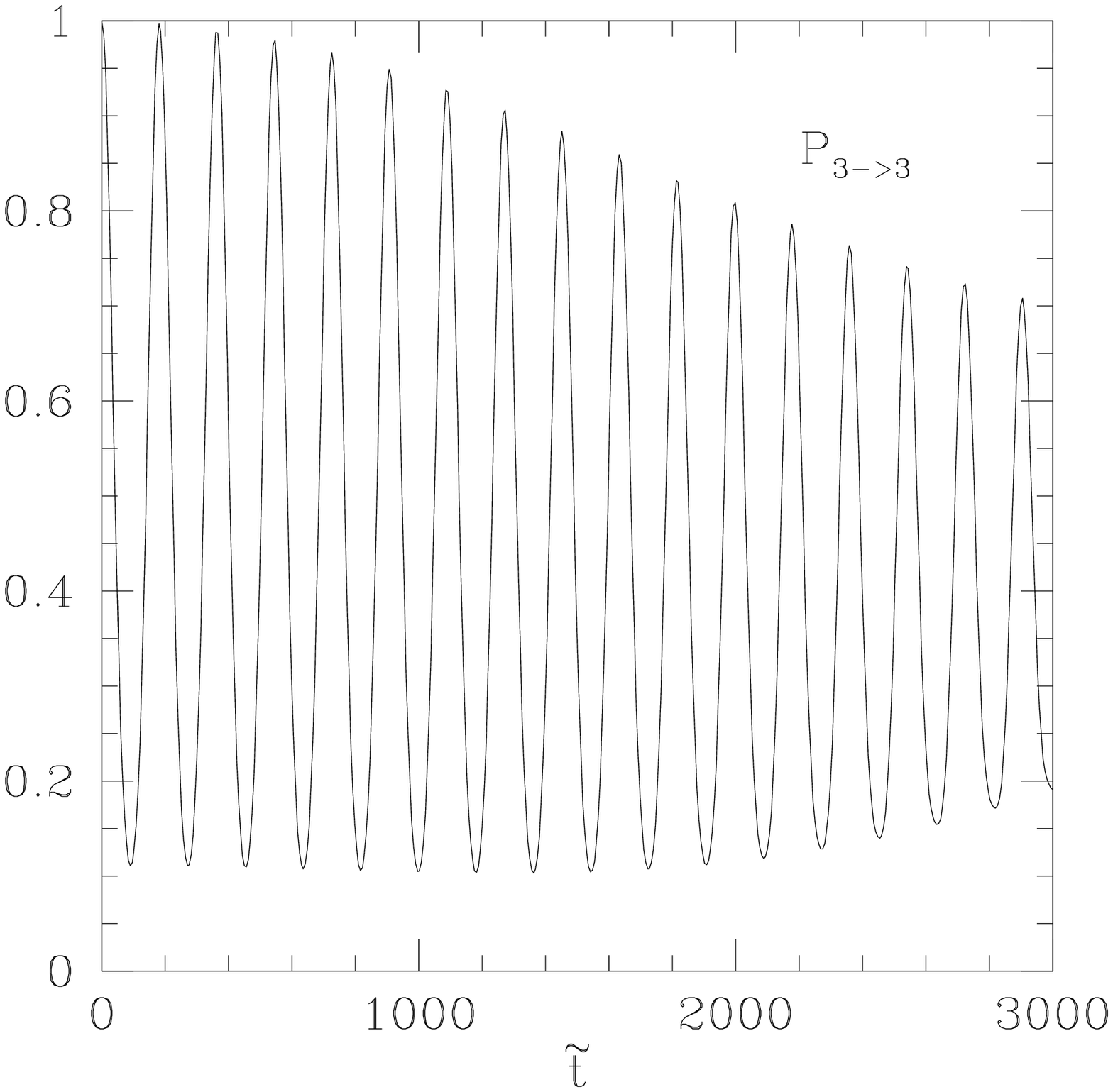}
 }
\caption{ Bulk-mediated neutrino flavor oscillations with $m=0.01$,
       $m_2=1$, and $m_{1,3}= 1\mp \delta m/2$ for $\delta m=10^{-3}$. 
       Despite the rather complicated nature of these flavor oscillations,
         the brane/bulk coupling is small and all flavor mixing
         angles on the brane are exactly zero.
   }
\label{fourstate}
\end{figure}

When all three of 
the brane neutrino flavors
are approximately degenerate, so that 
$m_1\approx m_2\approx m_3 \approx k\in \IZ$, 
we find that 
the bulk neutrino mediates an effective four-state
flavor oscillation which closely resembles a three-state oscillation on the
brane.  
As an example, let us consider the case 
$m=0.01$, $m_2 =1$, and $m_{1,3}=1\mp \delta m/2$.
The resulting probabilities for flavor oscillation and preservation
are then shown in Fig.~\ref{fourstate}.
Although the probabilities for flavor preservation and
conversion are flavor-independent for short times (top
plot of Fig.~\ref{fourstate}), a strong flavor dependence
develops for later times (remaining plots).
In fact, the corresponding time-averaged probabilities 
are given exactly by
\beqn
         \overline{P_{1\to 1}} = 4/9 ~,&~~~~~~~
         \overline{P_{1\to 2}} = 5/18 ~,&~~~~~~~
         \overline{P_{1\to 3}} = 1/9 ~,\nonumber\\
         \overline{P_{2\to 2}} = 5/18 ~,&~~~~~~~
         \overline{P_{2\to 3}} = 5/18 ~,&~~~~~~~
         \overline{P_{3\to 3}} = 4/9~. 
\label{taprobs}
\eeqn
Thus, if we start with $\ket{\nu_2}$, we see that we are equally
likely to have conversion into each flavor, whereas if we start
with $\ket{\nu_{1}}$ or $\ket{\nu_3}$,  we are more likely to preserve 
the initial flavor, with equally spaced declining probabilities for 
conversion into nearby flavors.
However, no matter which flavor of brane neutrino
we begin with, we see that exactly 1/6 of the initial probability
is ultimately lost into bulk neutrinos.

The above discussion pertains to only several specific choices of
parameters $\lbrace m, m_1,m_2,m_3\rbrace$ in this model;
a more complete discussion of these and other scenarios will be
presented in Ref.~\cite{ds}. 
In all cases, however, we find
that we can generate sizable neutrino oscillations
even if the brane/bulk coupling parameter $m$ is extremely small
and the brane theory is flavor-diagonal.
In general, large brane/bulk oscillations are triggered when
any of the $m_i$ take values that are near integers (in units of 
$R^{-1}$).  
Note, however, that this does not require a large fine-tuning.
When the extra dimensions are large, the corresponding 
Kaluza-Klein states are relatively closely spaced.
Thus, regardless of the specific mass of a flavor neutrino on the brane,
there is always a Kaluza-Klein state in the bulk which is effectively degenerate
with it.  Thus, such brane/bulk resonances are natural and can therefore
mediate large brane/bulk neutrino oscillations.
On the other hand, when $m$ is small, 
sizable {\it flavor}\/ oscillations leading to flavor conversion 
still require approximate flavor degeneracies on the brane.

This situation changes dramatically 
as the brane/bulk coupling parameter $m$ 
becomes larger.  In such cases, a larger population of 
bulk Kaluza-Klein neutrino states
participates in the neutrino mixings, and it is no longer necessary 
to have flavor degeneracies on the brane in order to generate
sizable bulk-mediated flavor oscillations.

\begin{figure}[ht]
\centerline{ 
      \epsfxsize 3.0 truein \epsfbox {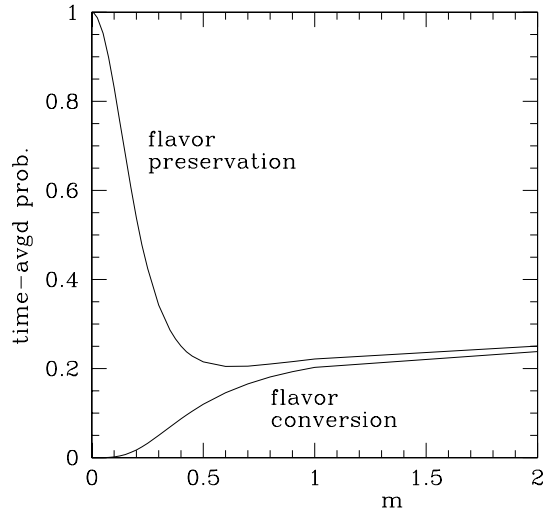}
 }
\caption{  
       Time-averaged probabilities for bulk-mediated flavor oscillations 
      as functions of the brane/bulk coupling $m$,
     with $m_1=1/2$, $m_2=3/2$, and $m_3 =20$.
        Although the flavor mixing angle on the brane remains zero,
    significant flavor conversion is generated for relatively small
     brane/bulk coupling $m$ even without flavor degeneracies on the brane.
   }
\label{newfig}
\end{figure}

As an example of this, let us consider the 
case when {\it none}\/ of the $m_i$ are close to an 
integer or to each other.
For example, we may take $m_1=1/2$, $m_2=3/2$, and $m_3 \gg 1$.
For small values of $m$, this situation gives neither
neutrino oscillations nor flavor conversion amongst the two
lightest flavors. 
However, as $m$ increases, 
both neutrino oscillations and flavor conversions are generated.  
This is shown in Fig.~\ref{newfig}, where we have plotted the time-averaged
probabilities for neutrino flavor preservation and conversion
as functions of the brane/bulk coupling $m$.

\begin{figure}[ht]
\centerline{ 
      \epsfxsize 2.0 truein \epsfbox {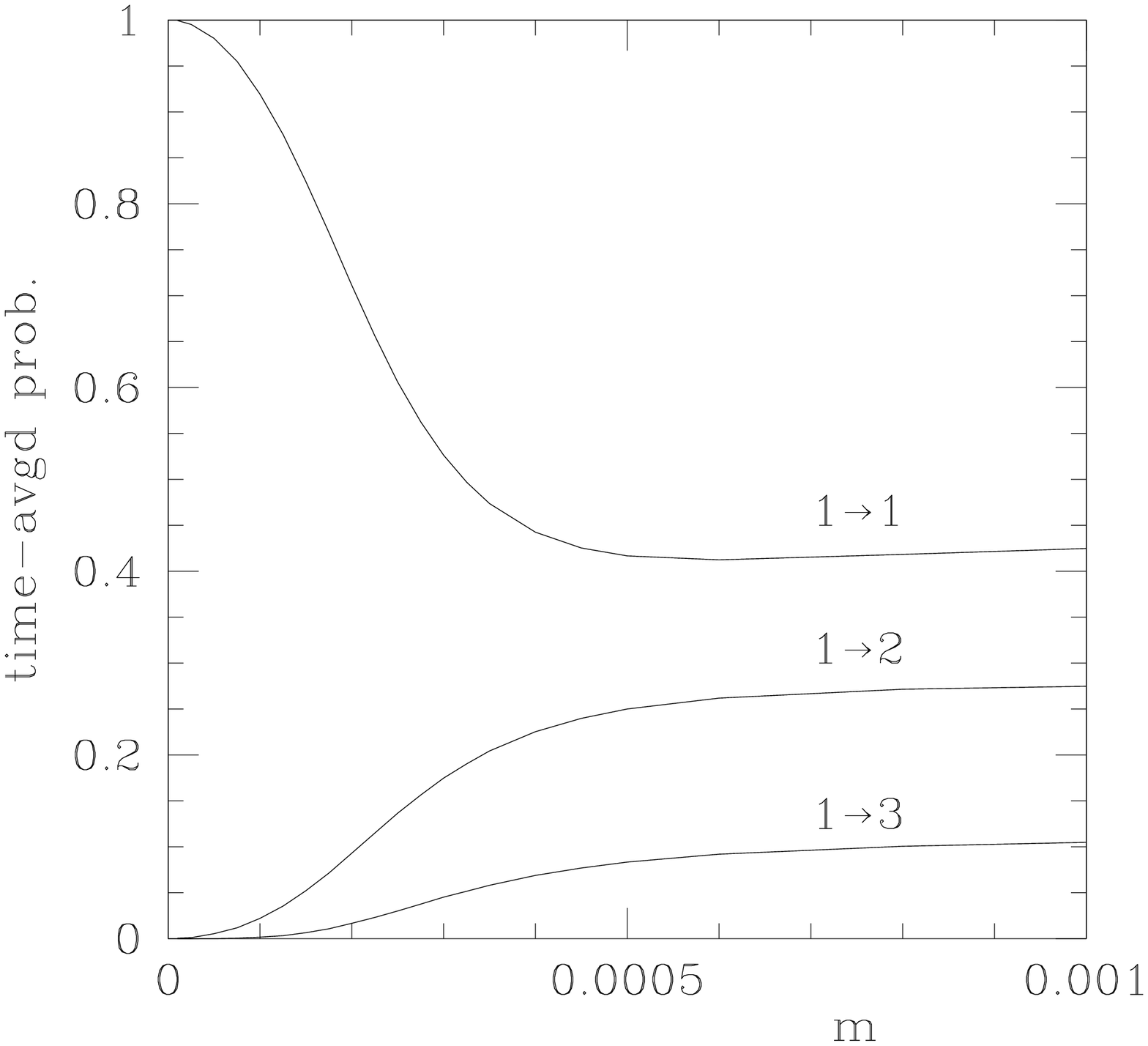}
      \hskip 0.2 truein
      \epsfxsize 2.0 truein \epsfbox {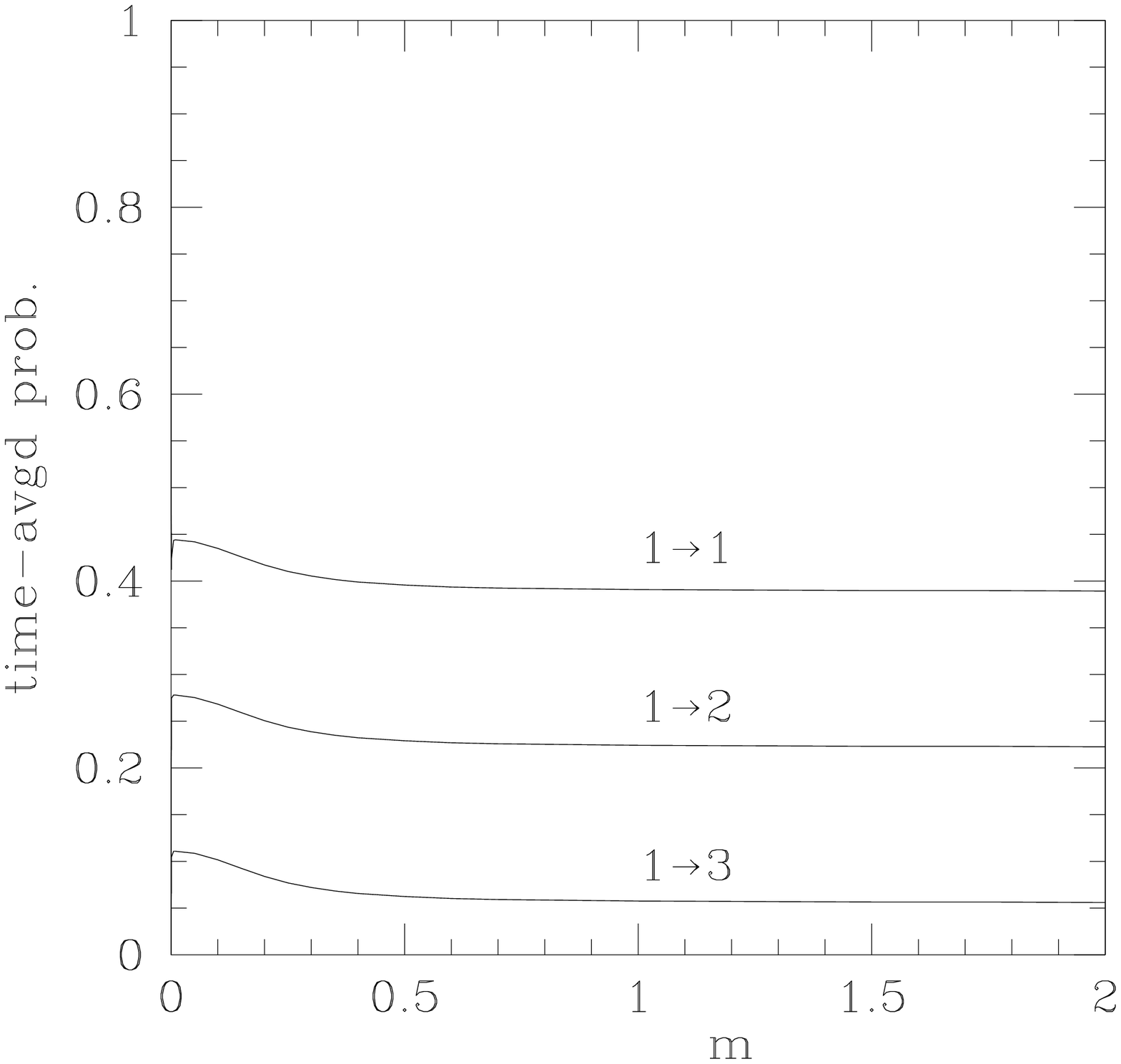}
 }
\centerline{ 
      \epsfxsize 2.0 truein \epsfbox {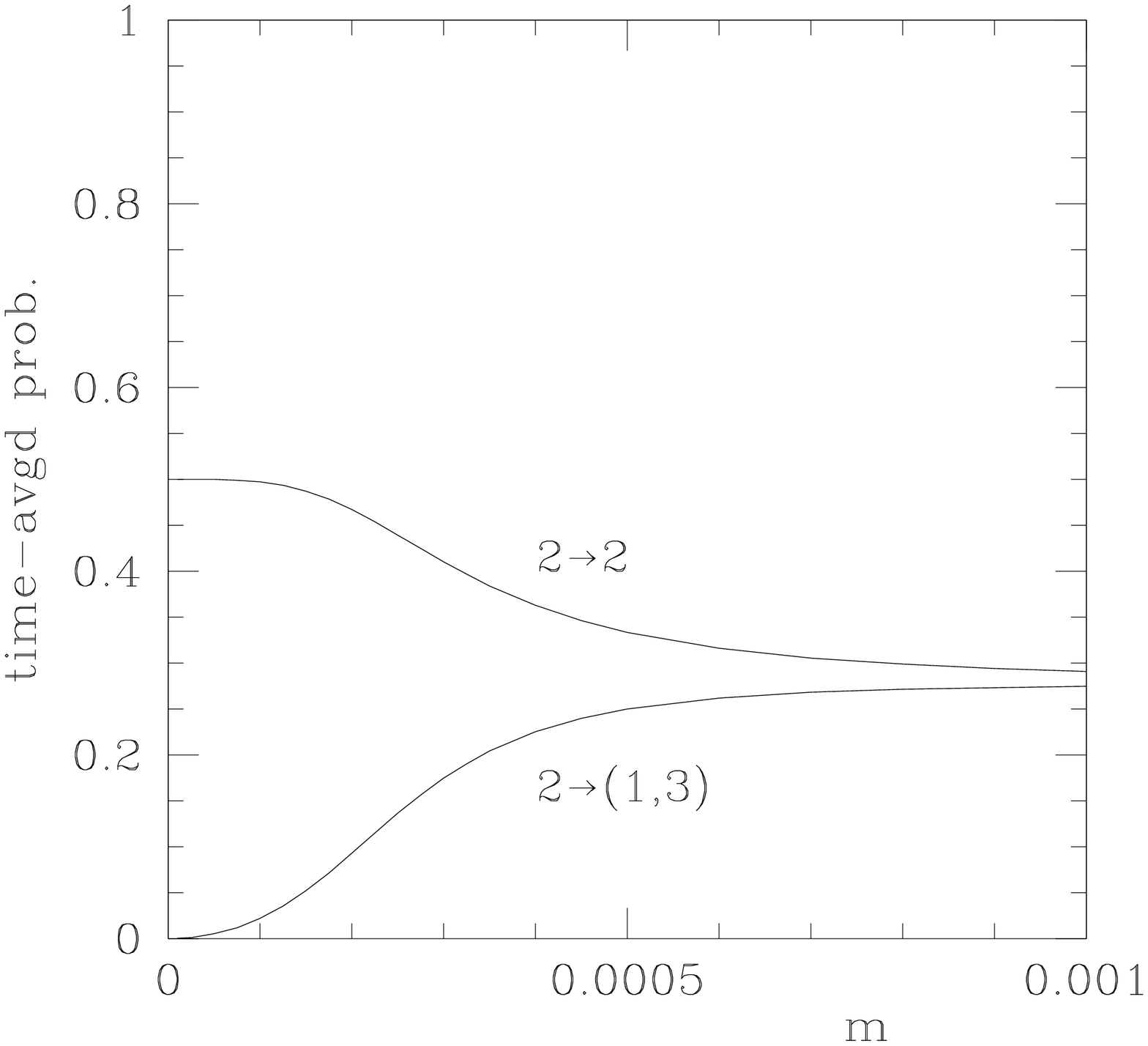}
      \hskip 0.2 truein
      \epsfxsize 2.0 truein \epsfbox {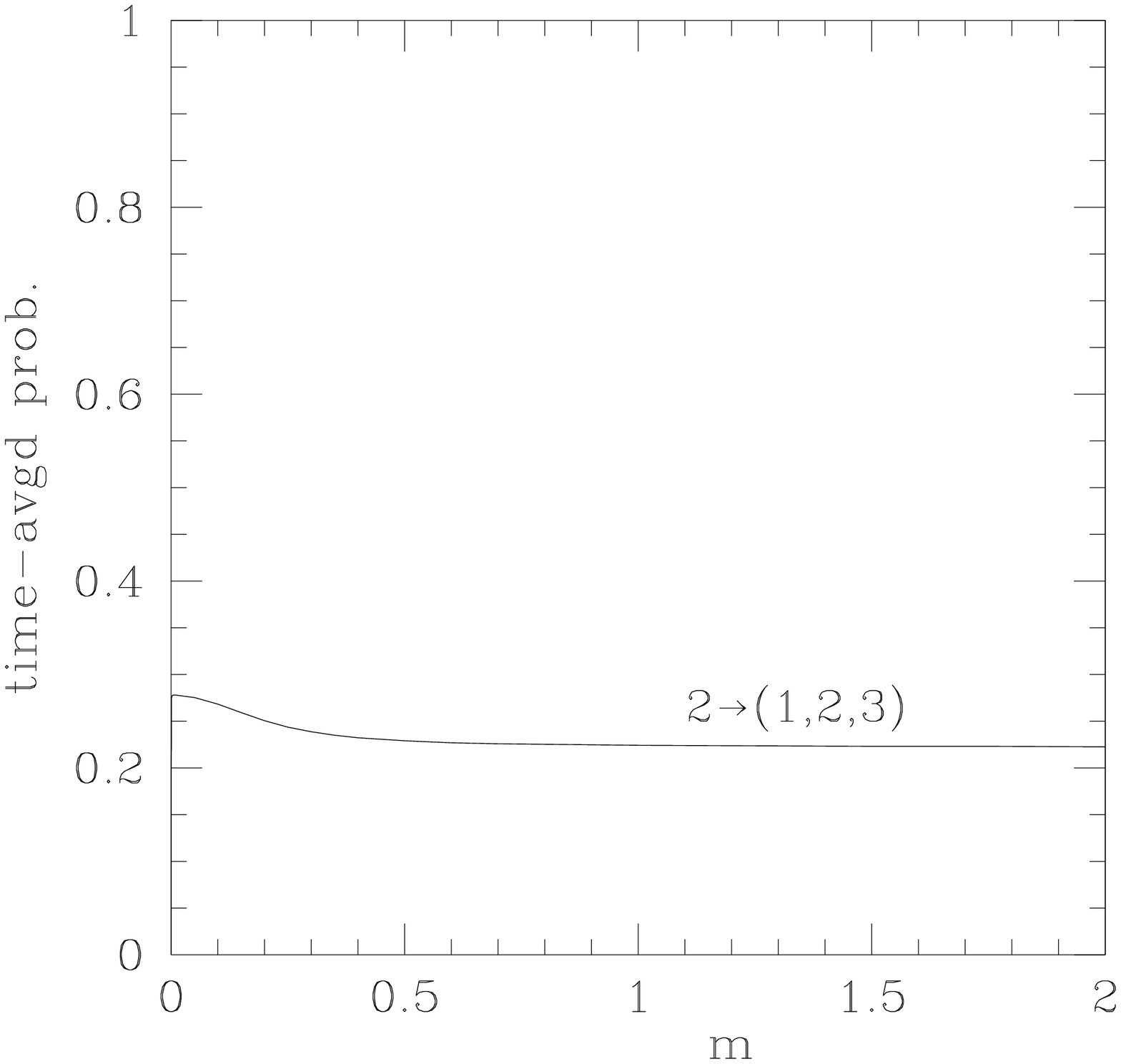}
 }
\caption{ Time-averaged probabilities for bulk-mediated neutrino
     flavor oscillations, as functions of the brane/bulk coupling $m$.
     We have taken $m_2=1$, $m_{1,3}= 1\mp \delta m/2$, and $\delta m=10^{-3}$.
     Note that with these parameters, $\overline{P_{3\to 3}}= \overline{P_{1\to 1}}$ 
      for all $m$.
   }
\label{tripleresonancefig}
\end{figure}

These effects can be further enhanced 
if any of the $m_i$ approaches an integer (in units of $R^{-1}$), 
for in such cases 
there is an
additional effect due to a strong resonance oscillation
between the corresponding brane neutrino and 
its degenerate Kaluza-Klein partner in the bulk.
This situation mirrors the phenomenon already discussed for the 
small-$m$ case.
As the most extreme case of this,
let us consider the situation in which all three of 
the brane neutrino flavors
are approximately degenerate with a bulk neutrino.
For concreteness we take $m_2=k \in \IZ$ 
and $m_{1,3}= k \mp \delta m/2$.
Choosing $k=1$ and $\delta m=10^{-3}$, 
we obtain 
the time-averaged probabilities shown in 
Fig.~\ref{tripleresonancefig}.

This figure may be understood as follows.
As $m\to 0$, 
we expect that the 
preservation probabilities should go to 1
while the conversion probabilities should go to zero.
This is indeed what happens, except that 
as $m\to 0$, the time-averaged probability $\overline{P_{2\to 2}}$
goes not to $1$ but to $1/2$.  This reflects the fact that for any non-zero
value of $m$, there continues to exist a maximal two-state resonance between
$\ket{\nu_2}$ and $\ket{N^{(1)}}$.  

As $m$ increases into the range $5\times 10^{-4}\lsim m \lsim 0.1$, 
we see from Fig.~\ref{tripleresonancefig}
that the neutrino probabilities then settle into an effective 
four-state resonance oscillation pattern
for which the time-averaged probabilities are given in (\ref{taprobs}).
As we see from Fig.~\ref{tripleresonancefig}, these time-averaged
resonance values persist over several orders of magnitude in $m$. 
Finally, as $m$ increases beyond $0.1$,
larger numbers of Kaluza-Klein states begin to 
participate in the neutrino oscillations.
This increases the probabilities for oscillations into bulk neutrinos,
which in turn causes our time-averaged neutrino probabilities on the brane
to decline relative to their values in (\ref{taprobs}).  
Remarkably, however, we see from Fig.~\ref{tripleresonancefig} that even though
these probabilities decline from the values given in (\ref{taprobs}),
they still preserve their relative spacings as functions of $m$.
Moreover,
we see that the time-averaged probabilities on the brane do not fall 
to zero as $m\to \infty$, 
but instead take the non-zero asymptotic values
\beqn
         \overline{P_{1\to 1}} = 7/18 ~,&~~~~~~~
         \overline{P_{1\to 2}} = 2/9 ~,&~~~~~~~
         \overline{P_{1\to 3}} = 1/18 ~,\nonumber\\
         \overline{P_{2\to 2}} = 2/9~,&~~~~~~~
         \overline{P_{2\to 3}} = 2/9 ~,&~~~~~~~
         \overline{P_{3\to 3}} = 7/18 ~.
\label{taprobsbigm}
\eeqn

The fact that these asymptotic probabilities are non-zero illustrates
that even when an {\it infinite}\/ number of bulk Kaluza-Klein states 
contribute to the neutrino oscillations, and even when the brane/bulk
coupling is {\it infinite}\/, we still do not lose
all of our initial neutrino probability into the bulk states! 
Instead, as $m \to \infty$,
we lose only 1/3 of the initial probability 
into bulk neutrinos. 
In other words, we see that
2/3 of the original probability remains ``on the brane'' even when
$m\to \infty$.
This is in fact a completely general feature of our model, 
and holds regardless of the specific values of the $m_i$.   
Moreover, this property generalizes:
with $n_f$ flavors on the brane, 
we find that only $1/n_f$ of the initial neutrino probability is lost into 
bulk neutinos.\footnote{  
      Note that this property relies crucially on the fact that we have
      taken only {\it one}\/ bulk neutrino in our
      model;  in other words, we have not extended flavor into the bulk. 
      By contrast, if we had taken a separate bulk neutrino for each brane neutrino, 
      we would have instead found that {\it all}\/ neutrino probability is ultimately
      lost into sterile bulk neutrinos as $m\to \infty$. 
     }
 
This surprising observation suggests that in  our model,
the large-$m$ case may 
be able to evade various four-dimensional bounds
(such as those from supernova cooling rates)
which ordinarily restrict the sizes of mixings with sterile neutrinos.
Indeed, our results such as those in Eq.~(\ref{taprobsbigm}) apply for 
large values of $m$ for which it is not possible to employ constraints derived  
using a ``perturbative'' approach in
which only the lightest few Kaluza-Klein bulk neutrinos mix significantly.
Rather, an exact treatment is required in which the full Kaluza-Klein tower
is incorporated simultaneously, for in our model it is the entire Kaluza-Klein tower 
that plays a role in reconverting bulk neutrinos back
into neutrino flavors on the brane as the brane/bulk coupling becomes large.
This issue, especially in conjunction with MSW matter effects,
will be discussed further in Ref.~\cite{ds}.

Thus, to summarize, we see that it is possible to have 
large effective ``bulk-mediated'' flavor mixing 
on the brane even when all flavor mixing angles on the brane
are zero.  
Moreover, this occurs even when the bulk theory is flavor-neutral
and the brane/bulk couplings are flavor-blind.
This illustrates that in higher dimensions, it may not be necessary
to have large flavor mixing angles on the brane in order  to accommodate
experimental data;  small (or even vanishing) flavor mixing angles on the  
brane may suffice.
Furthermore, even though this model contains only
five free parameters, it yields a surprisingly rich neutrino 
oscillation phenomenology, and exploits the higher-dimensional
nature of the bulk in an essential way.
Moreover, its essential structure involving only one bulk neutrino
leads to a radically different neutrino phenomenology for large
brane/bulk coupling than is possible in models with multiple
bulk neutrinos, and suggests that this limit might actually evade
supernova cooling constraints.
Thus, this model might be profitably used as the
basis of a detailed investigation of the experimental viability
of various higher-dimensional neutrino oscillation mechanisms.
This will be discussed further in Ref.~\cite{ds}.

\bigskip
\medskip
\leftline{\large\bf Acknowledgments}
\medskip

We wish to thank C.P.~Burgess,
B.~Campbell, L.~Dixon, Y.~Grossman, S.~Lola, 
R.~Mohapatra, K.~Orginos, and M.H.~Reno 
for discussions.
This work was supported in part by 
the National Science Foundation under Grant PHY-0071054
and by the Department of Energy under Grants
DE-FG02-95ER40906 and DE-FG03-93ER40792.
We also wish to acknowledge the hospitality of the
Aspen Center for Physics and the CERN Theory Division
where portions of this work were completed. 


\bigskip
\medskip

\bibliographystyle{unsrt}

\end{document}